\documentclass[twocolumn]{article} 
\usepackage[utf8]{inputenc}
\usepackage{amsmath}
\usepackage{graphicx}
\usepackage[a4paper, total={6in, 8in}]{geometry}
\usepackage{listings}
\usepackage{float}
\usepackage{placeins}
\usepackage[nottoc,numbib]{tocbibind}
\usepackage{xcolor}
\usepackage{hyperref}
\hypersetup{colorlinks,linkcolor={blue},citecolor={blue},urlcolor={blue}}
\usepackage[]{lineno}
\usepackage{bm}
\usepackage{url}
\usepackage[affil-it]{authblk}

\title{Comparison between electrostatic PP and PIC simulations on electron bunch expansion}

\author[1,2]{Yanan Zhang}
\author[2]{Xiaochun Ma}
\author[1]{Hui Liu}
\author[1,*]{Yinjian Zhao}

\affil[1]{School of Energy Science and Engineering,
Harbin Institute of Technology,
Harbin 150001, People’s Republic of China}
\affil[2]{Harbin Boiler Company Limited,
Harbin 150040, People’s Republic of China}
\affil[*]{Corresponding author: Yinjian Zhao, zhaoyinjian@hit.edu.cn}

\date{\today}

\begin{document}


\twocolumn[
  \begin{@twocolumnfalse}
    \maketitle
    \begin{abstract}

With the great development of parallel
computing techniques, the particle-particle (PP)
model has been successfully
applied in a number of plasma applications.
Comparing to particle-mesh (PM) models,
for example the widely used
particle-in-cell (PIC) method,
PP has the advantages of
high accuracy in solving Coulomb interactions.
In this paper, it is shown that
PP is also advantageous to
simulate non-neutral plasmas,
such as electron bunch expansion in vacuum.
The numerical effects of the macro-particle weight
and the time step length are investigated for
a PP model,
accurate and convergent results can be obtained
with less effort.
On the contrary,
PIC needs to simulate the same problem
with extremely large effort.
It is found that
the simulation accuracy does not
grow with reduced cell size monotonously,
thus no convergence can be easily obtained.
In the long run,
PIC must apply large enough domain to
cover all the expanding particles
and avoid non-physical effects caused
by imperfect infinite boundary condition,
which may result in too heavy computation
and make PIC infeasible.

    \end{abstract}
  \end{@twocolumnfalse}
]


\section{Introduction}

Plasma simulations using particles have been widely
applied in numerous areas to solve scientific and
engineering problems, such as nuclear fusion \cite{HIGGINSON2019439},
particle accelerator \cite{10.1063/5.0102919},
plasma propulsion \cite{Wang_2023},
space weather \cite{Gary_2018}, etc.
By modeling plasma particles (or macro-particles),
the equations of motion are solved
along with self-consistent fields and external fields,
such that particle trajectories and the
relevant physical process can be obtained.
In general, plasma simulation methods using particles
are more accurate than other approaches,
comparing to fluid models for example,
because the involved plasma in a particle model can be
in non-equilibrium state,
and those simulated particles
are able to describe any velocity distribution functions
in any transient states,
while fluid models are based on the assumption
of microscopic equilibrium.

According to the classification of
Hockney and Eastwood \cite{hockney},
particle simulation models can be
grouped into three broad categories,
the particle-particle (PP) model,
the particle-mesh (PM) model,
and the particle-particle-particle-mesh (PPPM)
model.
The idea of PP is simple and its implementation
is straightforward,
i.e., computation of
the forces between each pair of particles,
but two nested \textsf{do} loops
usually cannot be avoided,
which leads to huge computation ($\propto N_p^2$)
as the number of simulated particles $N_p$ increases.
Back to the time that the idea of particle
simulations was proposed,
the computers were way too slow to
carry out any meaningful PP simulations,
thus there were rare if not none PP related works
on solving plasma problems.
Then, the PM model was greatly developed,
such as the widely used particle-in-cell
(PIC) method
\cite{ck_birdsall_plasma_1991}.
Rather than solving inter-particle forces directly,
particles are deposited on mesh points
to obtain density values,
based on which the fields are solved
and interpolated back to particle positions.
The PM model can greatly reduce the computation
compared to the PP model
when many particles are simulated,
and the computation scales as
$\alpha N_p + \beta(N_c)$,
where $N_c$ is the number of cells applied
in one dimension,
$\alpha$ is a constant,
and $\beta$ is a function depending on the
field solver.
For example,
a typical PIC method solving the electrostatic
Poisson's equation
would require $\alpha=20$ and
$\beta = 5 N_c^3 \log_2 N_c^3$ \cite{hockney}.
However, the increase of computation speed of
PM is at the expense of the accuracy,
the forces within the range of
a cell are neglected,
thus the short-range particle interactions
can not be resolved.
Therefore, the PPPM model was proposed,
combining
PP to only solve the short-range interactions
and PM to reduce the computational cost,
which is more often applied in
molecular-dynamics simulations
\cite{PhysRevLett.101.135001}.

With the great development of
parallel computing techniques,
the PP model has been picked up
in recent years,
and a number of successful
applications have been carried out
for solving different plasma physics problems.
For example, because the PP model is good at
solving Coulomb collisions with the highest accuracy,
it has been applied in high-density collisional
plasmas to solve relaxation problems
in inertial confinement fusion \cite{ZHAO20172944},
and it can be used as benchmark to improve
Monte Carlo Coulomb collision models \cite{10.1063/1.5025581}.
In these applications, no macro-particles are used,
namely real electrons and ions are simulated
within a small simulation region, such that
real Coulomb collisions can be captured
by detailed particle trajectories.
There are also other areas that do
not have that many of charged particles
in the considered system,
such as in the scenario of plasma propulsion,
the plume of electrospray thrusters
is made of large charged droplets,
where PP is suitable to be applied
\cite{zhao,BREDDAN2023106079}.
By contract,
the problem is hard to be simulated by PIC,
because the particles have the same charge,
and strong Coulomb forces near the cone jet
cause the plume to expand quickly from micrometers to centimeters,
thus the mesh resolution of PIC
is hard to be coupled in this multi-scale problem.

In this paper, we present another physical problem that
is extremely suitable for PP, but very hard for PIC,
the electron bunch expansion in vacuum.
Such scenario is related to
applications of electron pulses for atomic diffraction
\cite{nature}
for example,
in which understanding of the electron
dynamics of Coulomb spreading is critical
\cite{JETP}.
First of all, the system is non-neutral,
since only electrons exist,
but PIC was originally designed for quasi-neutral plasmas.
For example,
the cell size criteria is usually determined by the Debye length,
which reflects the Debye screening distance only
for quasi-neutral plasmas.
From the perspective of the PIC algorithm, however,
macro-particles of electrons can still deposit their charges on the
grid points to obtain the charge density,
and the Poisson's equation can still be solved
as usual, but it would be difficult for PIC
to resolve the forces between particles within a cell.
Therefore, there seem to be a trade-off between the accuracy of solving the fields
and the computational cost.
However, we will show in this paper that
using more number of cells may not result in monotonously
increasing accuracy, and a convergence could not be achieved.
Furthermore, for three dimensional problems,
the computational cost is huge for PIC, because
the domain size has to be large enough to cover all expanding particles
and avoid
the non-physical effects caused by imperfect
infinite boundary condition.
On the contrary, the PP model can simulate
an infinite vacuum space straightforwardly,
by involving no charges other than the electron bunch.

The numerical effects of the PP model
on electron bunch expansion or other similar
problems with non-neutral plasmas,
have not been studied in detail.
Therefore, in this paper,
an electron bunch expansion in vacuum is considered,
being a representative example of non-neutral plasmas,
and the PP and PIC models are both applied and compared
to simulate the same problem.
Numerical effects of the macro-particle weight
and the time step length in PP are studied,
and numerical effects
of the mesh resolution, the domain size,
and the macro-particle shape factor in PIC
are investigated.
The paper is organized as follows.
In Sec.\ref{sec:2},
the PP algorithm, setup, and
simulation results are given.
The corresponding PIC simulations
are presented in Sec.\ref{sec:3}.
At last, conclusions are drawn in Sec.\ref{sec:4}.

\section{The PP Simulation}
\label{sec:2}

\subsection{The PP Algorithm and Setup}

Initially, an electron bunch is represented
by a number of macro-particles ${N_p}$.
Each macro-particle represents ${w_p}$ real particles.
It is assumed that all macro-particles share the same weight
for simplicity.
Giving the electron bunch a certain
spatial and velocity distribution initially,
and putting it in a vacuum,
such that the only forces acting on each electron
are due to other electrons,
the electron bunch will expand over time
due to the repulsive Coulomb forces.
The Coulomb's law reads
\begin{equation}\label{eq:A}
    \bm{F}_{ij}=\dfrac{1}{4\pi\varepsilon_0}\dfrac{Q_iQ_j}{\left|\bm{r}_{ij}\right|^2}\hat{\bm{r}}_{ij}
\end{equation}
where $\bm{F}_{ij}$ is the Coulomb force acting on the $i^{th}$
macro-particle due to the $j^{th}$ macro-particle,
$\bm{r}_{ij}=\bm{r}_i-\bm{r}_j$ is the position vector
pointing from the $i^{th}$ macro-particle to the $j^{th}$
macro-particle, $\hat{\bm{r}}_{ij}=\bm{r}_{ij}/
\left|\bm{r}_{ij}\right|$
denotes the unit vector,
$Q$ denotes the charge of the macro-particle
and $\varepsilon_0$ is the vacuum permittivity.
Then, the force acting on the $i^{th}$ macro-particle
due to all other macro-particles can be computed
through a sum,
\begin{equation}\label{eq:B}
    \bm{F}_{i}=Q_i\bm{E}_i=\dfrac{Q_i}{4\pi\varepsilon_0}\sum_{j=1,j\neq{i}}^{N_p}\dfrac{Q_j}{\left|\bm{r}_{ij}\right|^2}\hat{\bm{r}}_{ij}
\end{equation}
where $\bm{E}_i$ is the electric field at the position of the
$i^{th}$ macro-particle due to all other macro-particles.
Eq.(\ref{eq:B}) can be computed for all macro-particles, then the
(non-relativistic) equations of motion can be used to numerically
integrate or update the macro-particle velocities and positions,
\begin{equation}\label{eq:C}
    \dfrac{d\bm{r}_i}{dt}=\bm{v}_i
\end{equation}
\begin{equation}\label{eq:D}
    \dfrac{d\bm{v}_i}{dt}=\dfrac{\bm{F}_i}{M_i}=\dfrac{Q_i\bm{E}_i}{M_i}=\dfrac{q_i\bm{E}_i}{m_i}=\bm{a}_i
\end{equation}
where $\bm{v}_i$ denotes the velocity, $\bm{a}_i $ denotes the acceleration, $M_i$ denotes the mass of the $i^{th}$ macro-particle, and note that if a real particle has charge $q$ and mass $m$, the macro-particle charge is $Q=qw_p$ and mass is $M=mw_p$, such that the macro-particle weight can be canceled in Eq.(\ref{eq:D}). Furthermore, since we are dealing with electrons here with charge $q_e=-e$ and mass $m_e$, we can simplify Eq.(\ref{eq:B}), Eq.(\ref{eq:C}) and Eq.(\ref{eq:D}) to 
\begin{equation}\label{eq:E}
    \bm{F}_{i}=q_ew_p\bm{E}_i=\dfrac{{q_e}^2{w_p}^2}{4\pi\varepsilon_0}\sum_{j=1,j\neq{i}}^{N_p}\dfrac{\hat{\bm{r}}_{ij}}{\left|\bm{r}_{ij}\right|^2}
\end{equation}
\begin{equation}\label{eq:F}
    \dfrac{d^2\bm{r}_i}{dt^2}=\dfrac{d\bm{v}_i}{dt}=\dfrac{\bm{F}_i}{w_pm_e}=\dfrac{q_e}{m_e}\bm{E}_i=\bm{a}_i
\end{equation}
which are the governing equations that close the system for solving the electron bunch expansion.

\begin{figure*}[h!]
\centering
\includegraphics[width=0.35\textwidth]{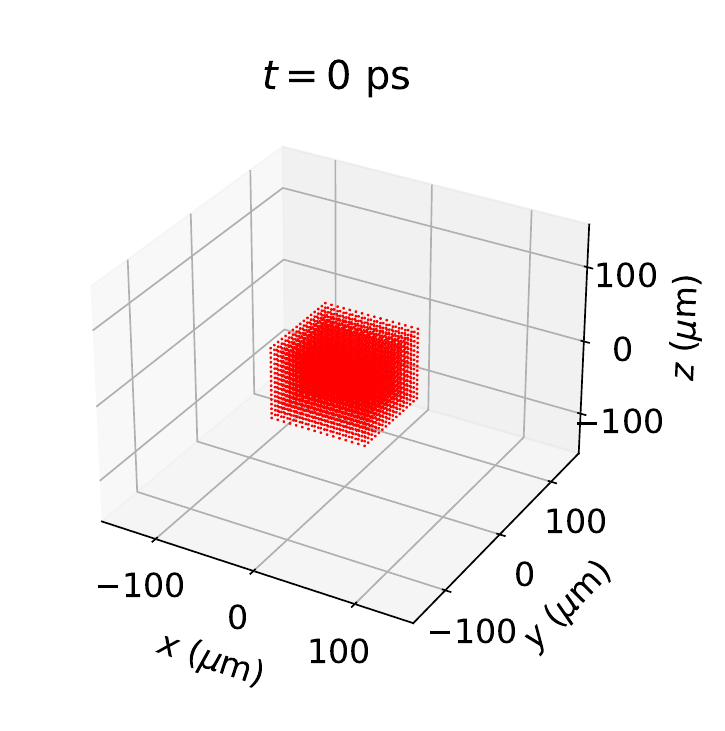}
\includegraphics[width=0.35\textwidth]{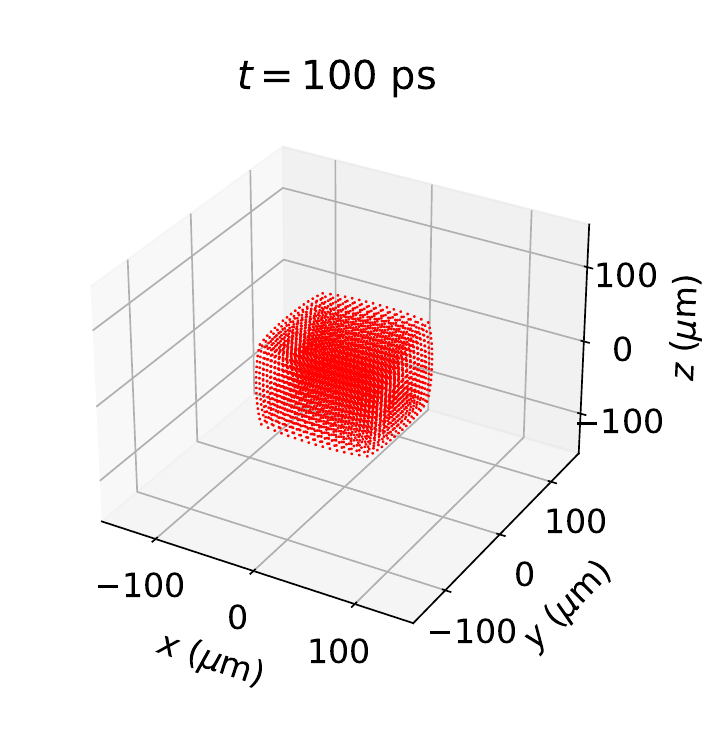}\\
\includegraphics[width=0.35\textwidth]{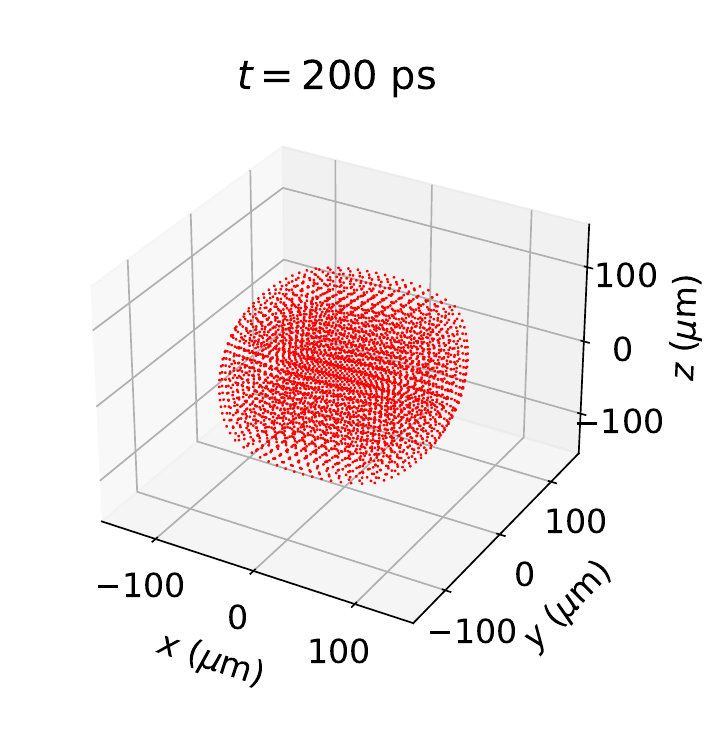}
\includegraphics[width=0.35\textwidth]{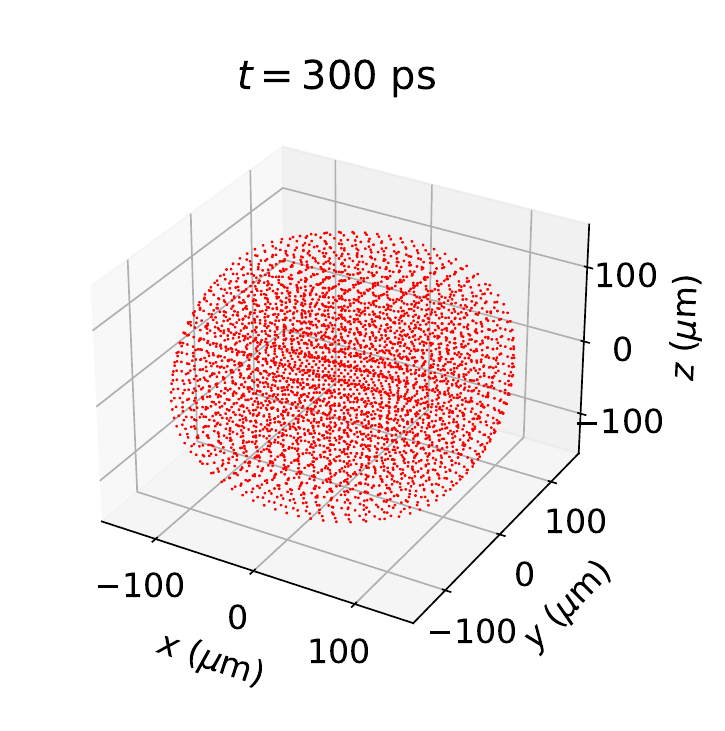}
\caption{
Snapshots at three different times of
3D scatter plots of electron bunch macro-particles.
}
\label{fig:xyz}
\end{figure*}

Because the computation of Eq.(\ref{eq:E})
for all particles
is known to be huge (scale as ${N^2_p}$), we apply MPI and OpenMP parallelization right away on the algorithm.
The whole particle array
containing the positions, velocities,
and accelerations of all particles is splitted by
different MPI ranks,
such that each MPI rank only needs to address
its own portion.
An additional particle array is used to
store the positions of all particles,
and the \textsf{mpi\_allgather} function
is used to gather each MPI's particle positions
at every time step after the positions are updated.
Within each MPI's two nested loops on solving
the Coulomb's field,
OpenMP is used to further parallelize the code.

Next, we apply the Velocity Verlet method
\cite{10.1063/1.442716} on numerically integrating the equations of motion Eq.(\ref{eq:F}). The Velocity Verlet method can be summarized as follows. At time step t, we have the position $\bm{r}(t)$, the velocity $\bm{v}(t)$, and the acceleration $\bm{a}(t)$ evaluated based on Coulomb's law using $\bm{r}(t)$. Then , we can obtain the new position at time step $t+\Delta t$,
\begin{equation}\label{eq:G}
    \bm{r}(t+\Delta t)=\bm{r}+\bm{v}(t)\Delta t+\bm{a}(t)\Delta t^2/2
\end{equation}
where $\Delta t$ denotes the timestep. Next, we can evaluate $\bm{a}(t+\Delta t)$ from $\bm{r}(t+\Delta t)$ from the Coulomb's law, and obtain the new velocity at $t+\Delta t$,
\begin{equation}\label{eq:H}
    \bm{v}(t+\Delta t)=\bm{v}(t)+[\bm{a}(t)+\bm{a}(t+\Delta t) ]\Delta t/2
\end{equation}
Now, we have obtained $\bm{r}(t+\Delta t)$,
$\bm{v}(t+\Delta t)$,
and $\bm{a}(t+\Delta t)$,
such that the procedure can be
repeated to obtain quantities at
$t+2\Delta t$ and so forth.

The simulation initial setup is
shown in Fig.\ref{fig:xyz} at 0 time step,
where the electron bunch has a cubic shape
with exact uniform spatial distribution,
i.e., the same inter-particle distances.
The reason that a cubic shape is chosen is
that it is easier for comparison with
the PIC simulation later.
The cube size is $l=0.1$ mm,
the electron bunch number density is
$n_0 = 5 \times 10^{16}$ m$^{-3}$,
and the number of macro-particles applied
is $N_p = 16 \times 16 \times 16 = 4096$,
which results in a macro-particle weight
$w_p = n_0 l^3 / N_p \approx 12.2$.
A relatively small timestep is chosen
$\Delta t = 10^{-13}$ s $= 0.1$ ps.
For simplicity, the initial velocities
of all macro-particles are set to be zero.

\subsection{The PP Results}

The simulation results are presented in this section. 
First, snapshots at four different times
are shown in Fig.\ref{fig:xyz}, which are the macro-particle
3D scatter plots.
We can see that due
to the Coulomb repulsion, the electron bunch expands overtime.
This simulation takes about 532 s to finish 100000
time steps using 64 OpenMP threads and 1 MPI rank on a computer with two AMD EPYC 9174F CPUs.

\begin{figure}[h!]
\centering
(a)
\includegraphics[width=0.45\textwidth]{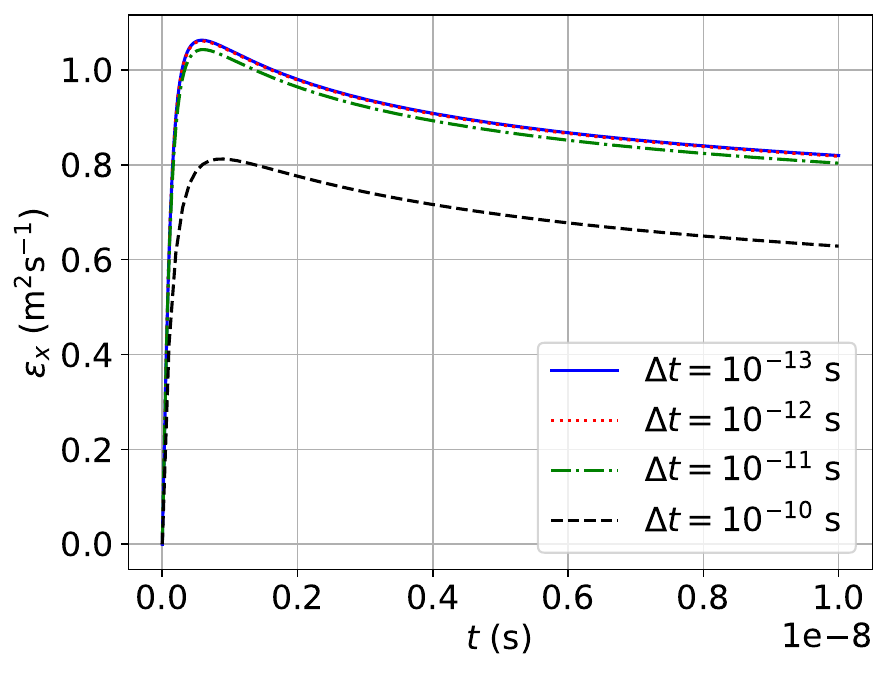}
(b)
\includegraphics[width=0.45\textwidth]{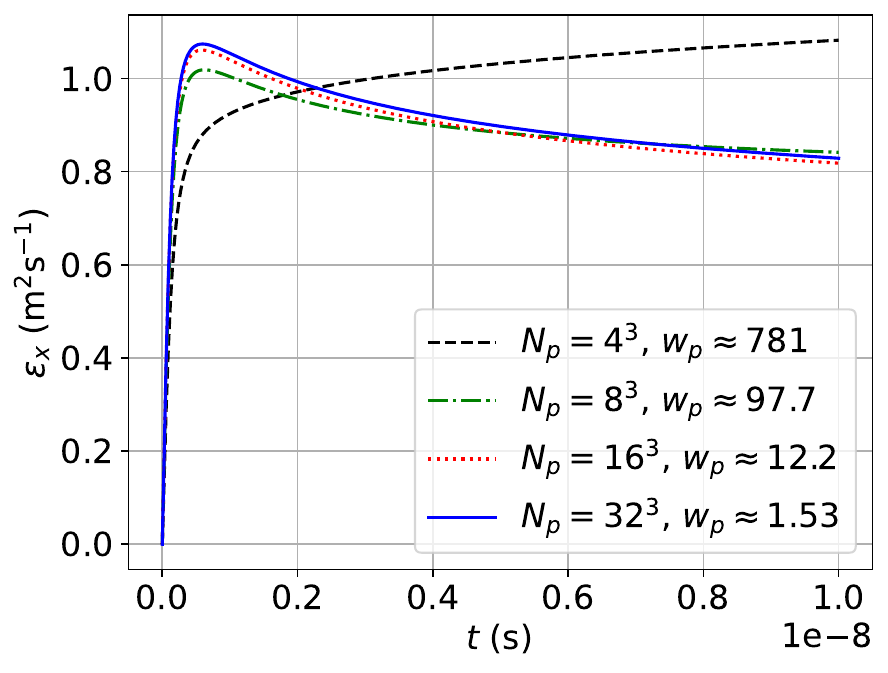}
\caption{Emittance over time for cases with
(a) different timesteps and
(b) different number of macro-particles.
}
\label{fig:em}
\end{figure}

To verify the validity of the PP results,
we first show the influence of the timestep $\Delta t$,
which is reduced from $10^{-13}$ s to $10^{-10}$ s.
The emittance $\varepsilon_x$ over time
is plotted for four cases with different $\Delta t$
in Fig.\ref{fig:em} (a).
Note that the emittance is defined as
\begin{equation}\label{eq:I}
    \varepsilon_x=\sqrt{{\langle}x^2{\rangle}{\langle}v^2_x{\rangle}-{\langle}xv_x{\rangle}^2},
\end{equation}
which is a statistical quantity to reflect
the beam quality in phase space
\cite{PhysRevSTAB.6.034202},
where the angle brackets represent
an average over all particles.
As we can see, $\Delta t=10^{-10}$ s is obviously
not small enough for the simulation,
while the other three smaller $\Delta t$ leads
to closer and more accurate results.
As $\Delta t$ increases 10 times,
the total number of time steps needed to finish $10^{-8}$ s
decreases 10 times,
thus the simulation can be finished 10 times faster.

Next, we show the influence of the number of macro-particles $N_p$
(or the macro-particle weight $w_p$).
Four cases are tested from $N_p = 4^3 = 64$ to $N_p = 32^3 = 32768$,
while $\Delta t = 10^{-13}$ s is still applied.
Note that for the case with $N_p = 32^3$,
the macro-particle weight $w_p \approx 1.53$ is already very close to 1,
meaning we are pretty much simulating real electrons.
The results are shown in Fig.\ref{fig:em} (b),
we can see that $N_p = 4^3$ is obviously inaccurate,
and as $N_p$ increases, the results become more accurate.
However, one should note that this physical problem we are dealing with
is difficult in terms of long-term accuracy, because the most important physics
happens at the very beginning of the simulation, where all electrons are placed
within a small region with strong Coulomb forces,
any small errors of computing the forces at the beginning will lead to growing errors
in macro-particle velocities, positions, and the beam emittance in the long run.


\begin{figure}[!ht]
\centering
\includegraphics[width=0.35\textwidth]{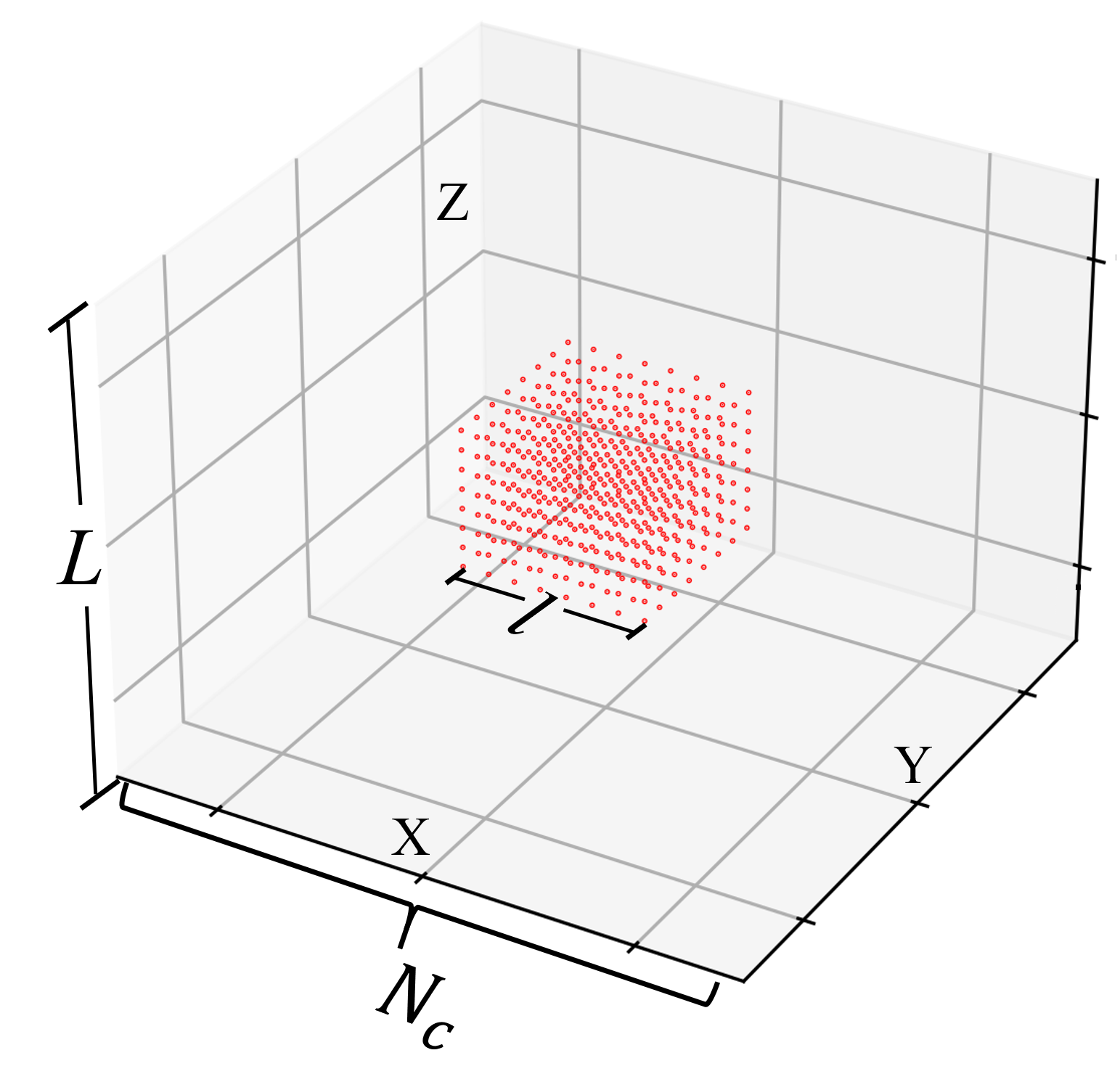}
\caption{The PIC simulation setup.
}
\label{fig:pic_setup}
\end{figure}

\begin{figure*}[!ht]
\centering
\includegraphics[width=0.7\textwidth]{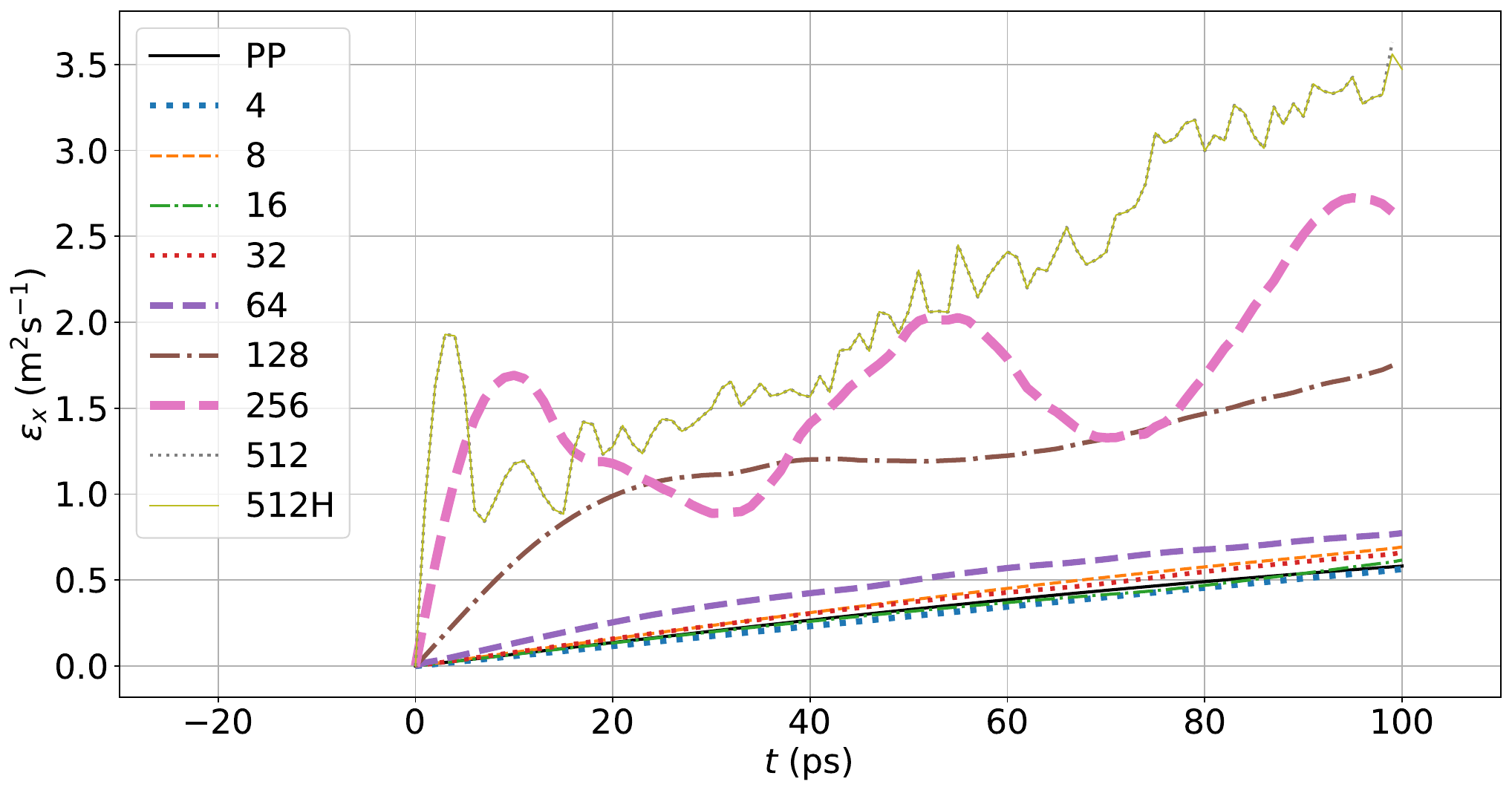}
\caption{
Comparison of emittance between PP and PIC with different mesh resolution.
The number of cells per direction $N_c$ is
labeled
for each curve.
}
\label{fig:warpx_em}
\end{figure*}

\begin{figure}[!ht]
\centering
\includegraphics[width=0.35\textwidth]{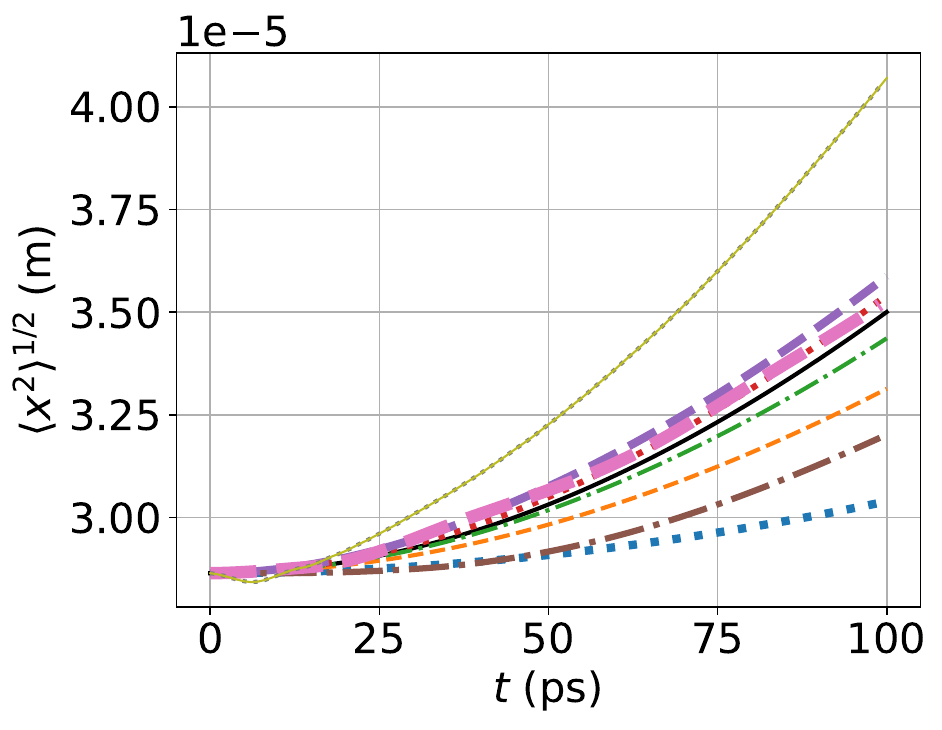}
\includegraphics[width=0.35\textwidth]{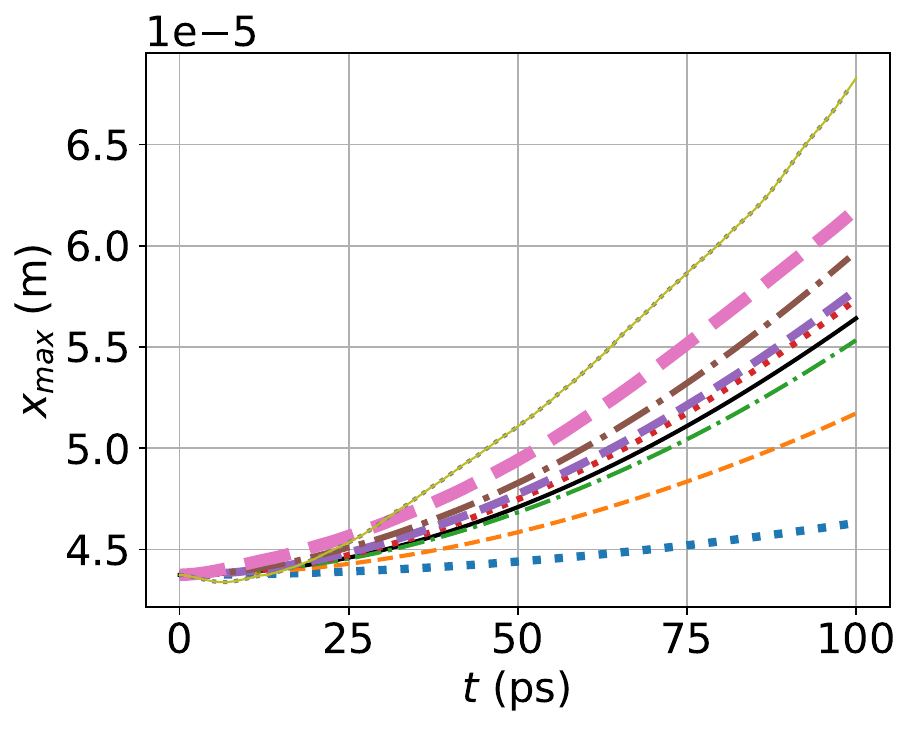}
\caption{
Comparison of
$\langle x^2 \rangle^{1/2}$
and $x_{max}$
between PP and PIC with different mesh resolution.
The line legend is the same as
the one in Fig.\ref{fig:warpx_em}
}
\label{fig:warpx_xm}
\end{figure}

\begin{figure}[!ht]
\centering
\includegraphics[width=0.35\textwidth]{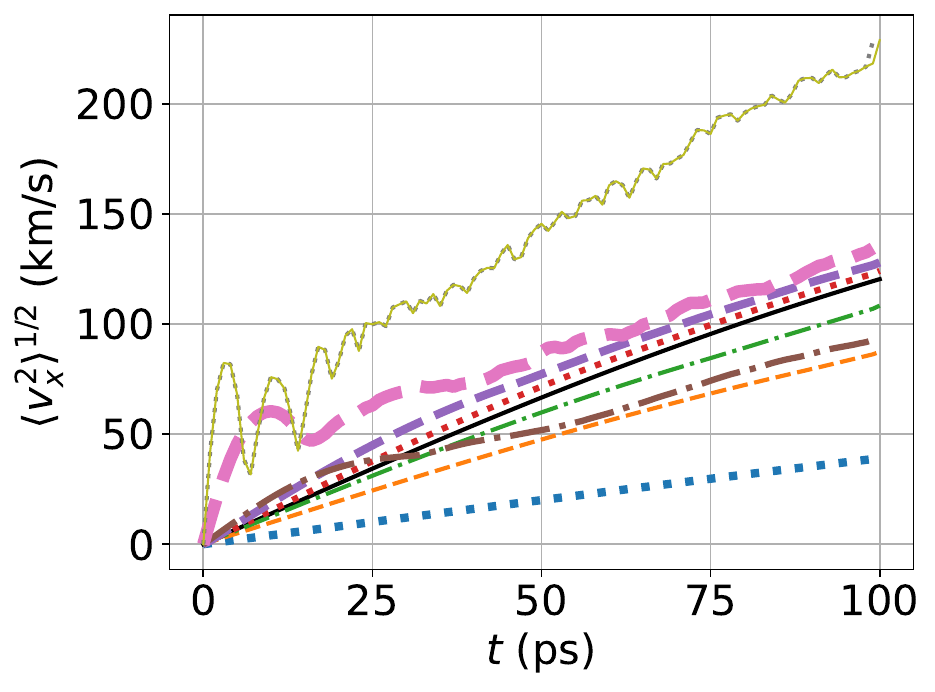}
\includegraphics[width=0.35\textwidth]{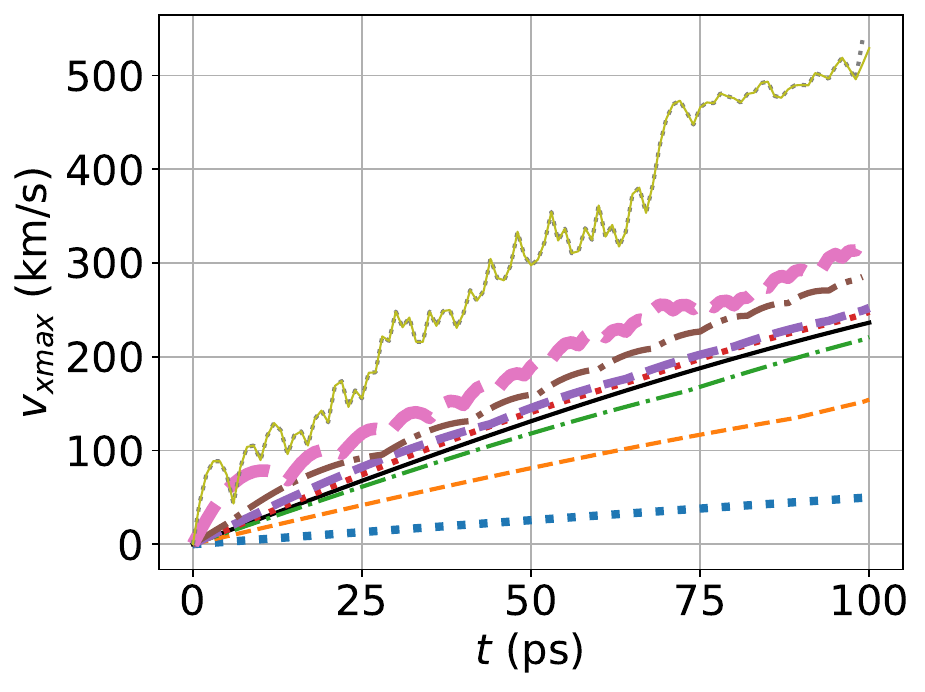}
\caption{
Comparison of
$\langle v_x^2 \rangle^{1/2}$
and $v_{xmax}$
between PP and PIC with different mesh resolution.
The line legend is the same as
the one in Fig.\ref{fig:warpx_em}
}
\label{fig:warpx_vxm}
\end{figure}

\begin{figure*}[!ht]
\centering
\includegraphics[width=0.35\textwidth]{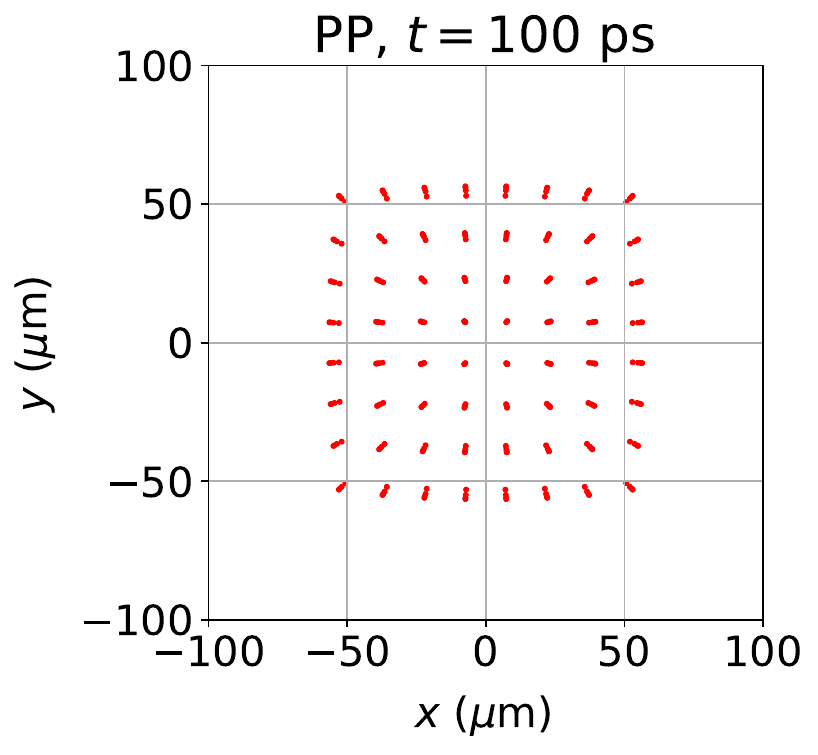}
\includegraphics[width=0.35\textwidth]{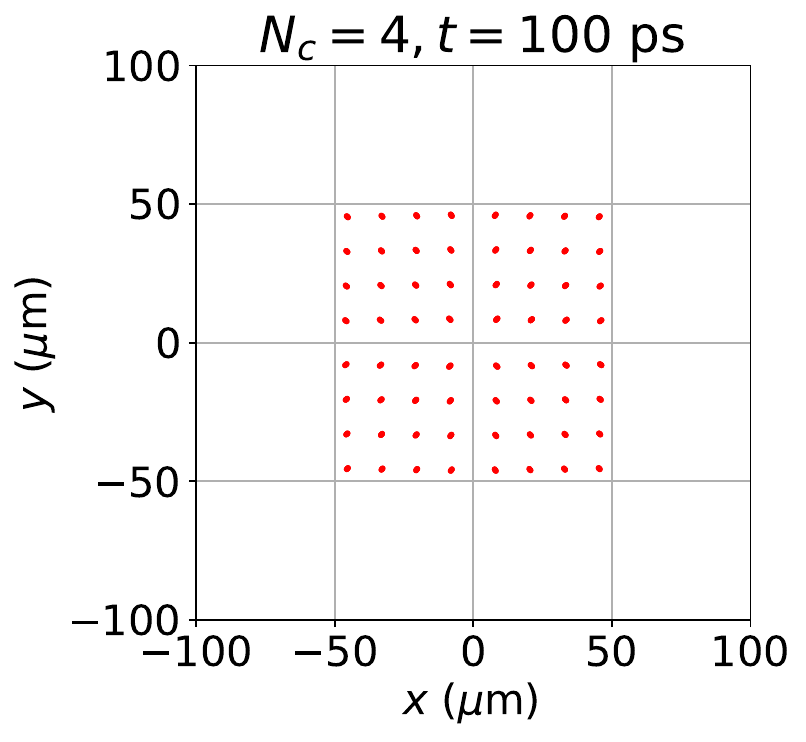}\\
\includegraphics[width=0.35\textwidth]{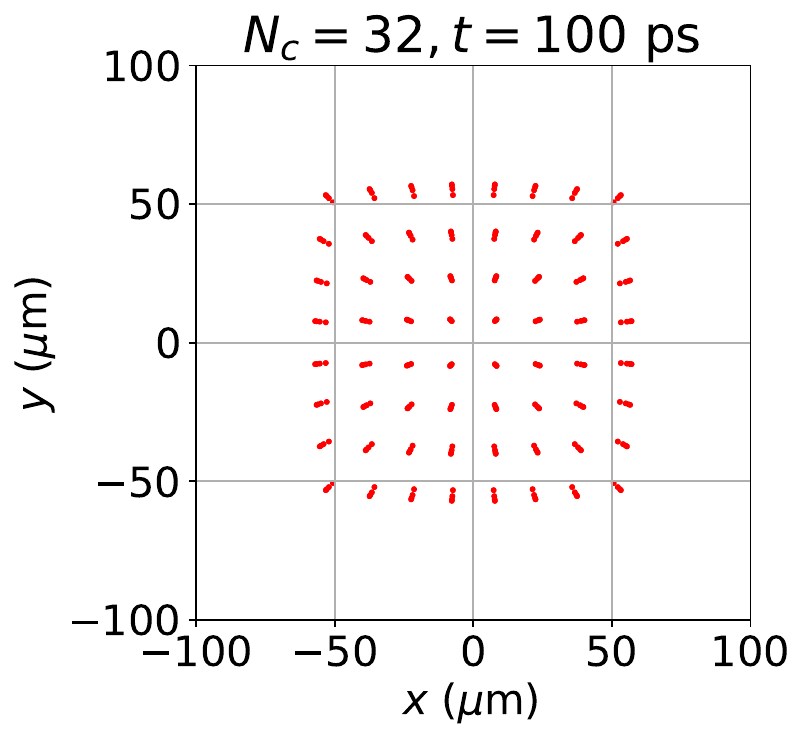}
\includegraphics[width=0.35\textwidth]{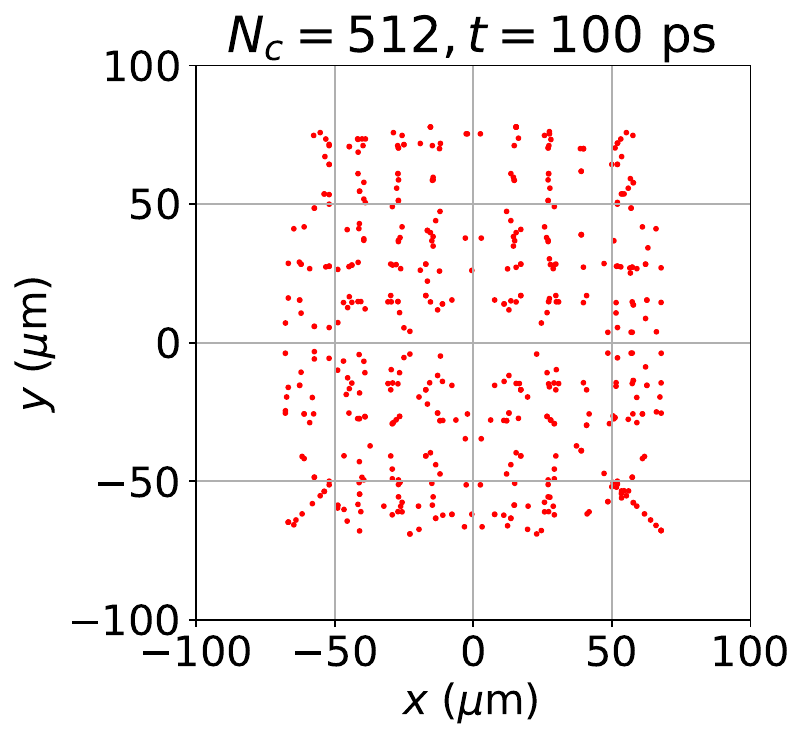}
\caption{
2D spatial plots 
of PP and PIC with different mesh resolution.
}
\label{fig:warpx_p}
\end{figure*}

\section{The PIC Simulation}
\label{sec:3}

\subsection{The PIC Algorithm and Setup}

In this section,
we simulate the electron bunch expansion using a standard electrostatic(ES) PIC method, and compare the results to those obtained using the PP model.
In an ES PIC, the Poisson's equation is solved based on the charge density on the grid deposited by macro-particles. Note that the electrostatic Poisson's equation is physically equivalent to Coulomb's law, but Poisson's equation is described from the perspective of statistical charge density, while the latter is described by point charges. The PIC code used here is the open-source PIC code WarpX \cite{warpx}.
WarpX is a highly-parallel and highly-optimized code,
which can run on GPUs and multi-core CPUs.
WarpX scales to the world’s largest supercomputers and
was awarded the 2022 ACM Gordon Bell Prize.
The features applied in this paper are
the 3D Cartesian geometry,
the MLMG (Multi-Level Multi-Grid)
electrostatic field solver,
the Boris leap-frog particle pusher,
and the MPI parallelization.

The PIC simulation domain is set to be a cube with size $L=0.2$ mm. The macro-particles of the electron bunch are initially placed within a smaller cube with size $l=0.1$ mm, as illustrated in Fig.\ref{fig:pic_setup}, the same as the preceding
PP simulation.
The electron bunch number density is still $n_0=5\times10^{16}$ m$^{-3}$ initially. The time step is set to be $\Delta t=10^{-12}$ s. The number of macro-particles is set to be $N_p=8^3=512$. 
The WarpX's MLMG Poisson solver is applied with required precision $10^{-6}$($10^{-9}$ is also tested for one case).
The linear density deposition and field interpolation schemes
are applied.
All faces of the cube are set to be Dirichlet boundary
conditions with zero fixed potentials for fields
and absorbing boundary conditions for macro-particles.

\subsection{The PIC Results
and Comparison to PP}

Keeping the above parameters fixed, the precision of solving the
Poisson's equation depends on the number of cells
per direction $N_c$, or the cell size $L/N_c$. Several $N_c$ are tested, ranging from 4 to $2^9=512$, namely the cell sizes are varied from $50$ $\mu$m to $0.390625$ $\mu$m. Note that the distance between two macro-particles is $l/8=12.5$ $\mu$m. A corresponding PP test is carried out too, served as an accurate baseline, using the same $N_p=512$ and $\Delta t = 10^{-12}$ s.
The results of emittance $\varepsilon_x$,
$\langle x^2 \rangle^{1/2}$,
and
$\langle v_x^2 \rangle^{1/2}$
are presented in
Fig.\ref{fig:warpx_em},
Fig.\ref{fig:warpx_xm},
and
Fig.\ref{fig:warpx_vxm},
respectively.

The emittance plot in the Fig.\ref{fig:warpx_em} shows that the emittances of cases with $N_c\leq32$ are closer to that of PP, while the emittance of cases with $N_c\textgreater 32$ are even worse. This result may be contrary to our first thought that larger $N_c$ with smaller cell size should lead to more accurate and gradually converged results. However, if we further look at the ${\langle}x^2{\rangle}^{1/2}$ plot in Fig.\ref{fig:warpx_xm}(top), we can see that the results of cases $N_c =256,64,32,16$ are closer to that of PP. The result of $N_c=128$ is surprisingly worse. In addition, the plot of the maximum value of $x$ among all macro-particles, $x_{max}$ is plotted in Fig.\ref{fig:warpx_xm} (bottom). We can see that this time the lines are in order, an increasing $N_c$ leads to larger $x_{max}$.
Next, look at the ${\langle}v_x^2{\rangle}^{1/2}$ plot in Fig.\ref{fig:warpx_vxm} (top), the order is similar to that in Fig.\ref{fig:warpx_xm} (top). The maximum value of $v_x$ among all macro-particles, $v_{xmax}$, is plotted in Fig.\ref{fig:warpx_vxm} (bottom), from which we can see again the lines are in order, an increasing $N_c$ leads to larger $V_{xmax}$.

The above results indicate that:
(1) The accuracy of PIC does not monotonically increase
as the cell size decreases.
(2) When the cell size is too small,
nonphysically large forces appear
causing the electron bunch to expand too fast.
(3) Because the tendency of the averaged quantities is different
from that of the maximum quantities,
the locations of macro-particles
in the cube or cell matters.
The above conclusions can also be confirmed
by looking at the 2D spatial plots
in Fig.\ref{fig:warpx_p}.
When $N_c$ is only 4,
the cell size is 50 $\mu$m,
bigger than the inter-particle position 12.5 $\mu$m,
such that the forces within a cell are weakened,
and the electron bunch expands slower.
When $N_c$ is 512,
the cell size is 0.390625 $\mu$m,
so each macro-particle occupies a single cell,
and there are about 32 cells between two macro-particles.
From the plot, we can see the electron bunch expands faster,
and loses its symmetry especially at the diagonals.
When $N_c$ is 32,
the results are very close to that of PP,
and a longer simulation will be conducted later
to see if they can match in the long run.

In addition, a higher required precision with $10^{-9}$ is
set for the Poisson solver in PIC
for the case $N_c=512$ labeled as 512H
in Fig.\ref{fig:warpx_em},
Fig.\ref{fig:warpx_xm},
and
Fig.\ref{fig:warpx_vxm},
since the results overlap with that of the 512 case,
the required precision $10^{-6}$ is sufficient.
In terms of the computation speed,
the most heavy run with $N_c=512$ needs to solve
the Poisson's equation on a grid system with
$512 \times 512 \times 512$ cells
for 100 time steps.
It takes about 148 s when using 64 MPI
with domain decomposition $8 \times 8 \times 8$
on a computing node with
two AMD EPYC 9754 CPUs.
By contract, the corresponding PP run
takes only about 0.28 s without parallelization.
Thus,
under the scenario of the electron bunch expansion,
PP is much more accurate and efficient than PIC.

\begin{figure}[!ht]
\centering
\includegraphics[width=0.4\textwidth]{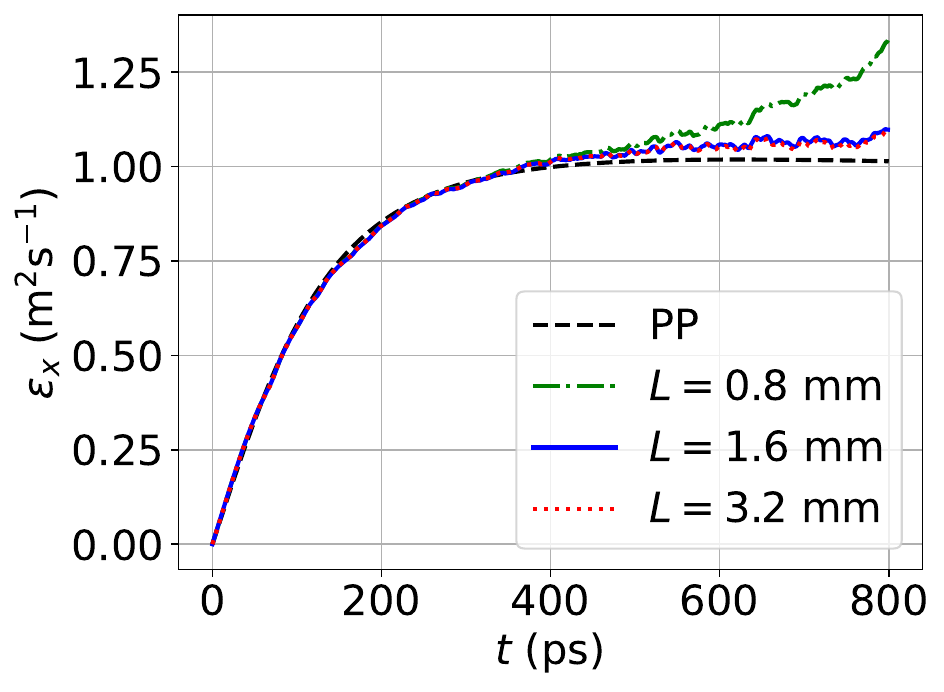}
\includegraphics[width=0.4\textwidth]{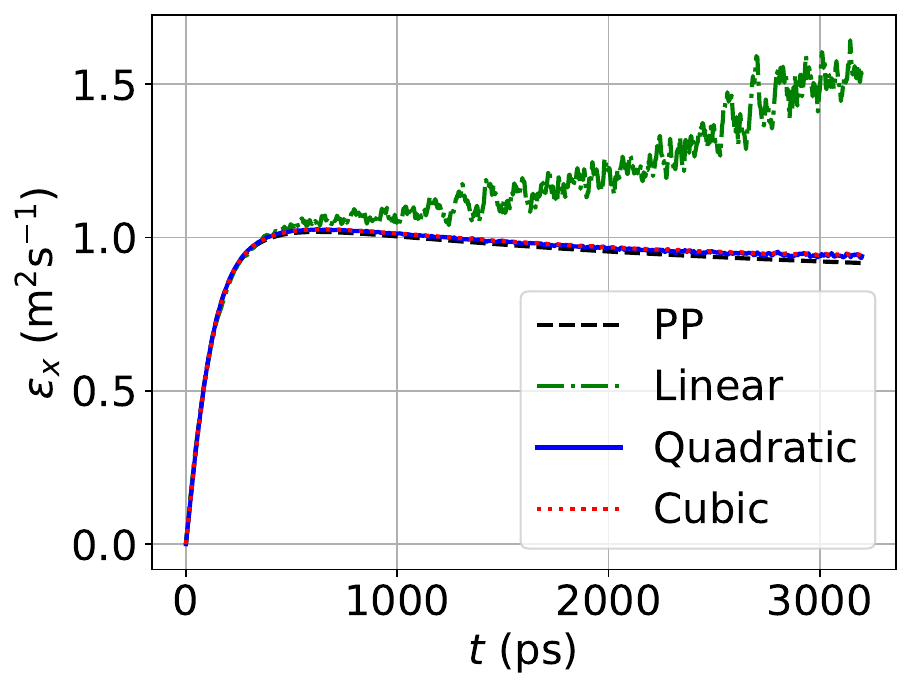}
\caption{
Comparison of emittance between PP and PIC in the long run
with varied domain size (top)
and varied particle shape-factor (bottom).
}
\label{fig:warpx_long}
\end{figure}

\subsection{Comparison between PIC and PP in the long run}

To further investigate if the $N_c=32$ case
with cell size 6.25 $\mu$m can match with
PP in the long run,
three tests are done with fixed cell size,
but enlarged $N_c$ and $L$ accordingly to make sure
the macro-particles do not move out of the domain
and evaluate the effects of the domain size in the meanwhile.
$N_c=128,256,512$ are considered
and
$L=0.8,1.6,3.2$ mm are varied accordingly.
The results of the emittance growth until
800 ps are plotted in Fig.\ref{fig:warpx_long} (top).
As we can see, all three cases match well with PP
at the beginning until about 400 ps.
Then the $L=0.8$ mm case
deviates the most from PP,
indicating the domain size is not large enough,
since a fixed potential Dirichlet boundary condition
is used, which is not a perfect infinite boundary.
The cases with $L=1.6$ and 3.2 mm overlap with each other,
indicating that further enlarging the domain size cannot
contribute to obtain more accurate results,
which would be due to the fact that
after the electron bunch expands,
eventually each macro-particle occupies a single cell,
and there are more and more cells between two
adjacent macro-particles,
so the fields cannot be solved accurately anymore,
just as shown already by the $N_c=512$ case
in Fig.\ref{fig:warpx_em},
Fig.\ref{fig:warpx_xm},
Fig.\ref{fig:warpx_vxm},
and
Fig.\ref{fig:warpx_p}.

At last,
another even longer run till 3200 ps is carried out
varying the particle shape factor
from linear, quadratic, to cubic.
To let the PIC simulation domain still
contains all particles till 3200 ps,
while maintaining the cell size unchanged,
$N_c=1024$ and $L=6.4$ mm have to be set,
resulting in a simulation time of about 37 hours
using 64 MPI ranks.
However, the corresponding PP run only
taks 23 s without parallelization.
The emittance results are shown in
Fig.\ref{fig:warpx_long} (bottom).
As we can see,
the linear case deviates from PP
obviously after about 800 ps,
while the quadratic and cubic
shape factors lead to only small deviation
at 3200 ps.
Thus higher-accuracy shape factors contribute
for PIC to obtain better results
in electron bunch expansion.

\section{Conclusion\label{sec:4}}

In this paper,
under the scenario of electron
bunch expansion in vacuum
due to Coulomb repulsion,
the PP model and the PIC model
are compared in detail.
For the PP model applied,
there are only two factors that
determine the accuracy,
one is the macro-particle weight,
the other is the time step length.
Simulation results indicate that
decreasing the macro-particle weight
and the time step length
leads to converged results,
and little effort is needed for
the PP model to conduct accurate simulations of
electron bunch expansion.
On the contrary,
it is very hard for PIC
to simulate electron bunch expansion,
because
the domain size needs to be large enough
to contain all expanding particles,
and the mesh resolution cannot be
too coarse and needs to be chosen carefully.
It is found that the relation between
the mesh resolution and the simulation accuracy
is not monotonous,
thus no convergence can be obtained.
When the cell size is close to the
initial inter-particle distance,
fairly accurate simulation results can be obtained.
The precision of the Poisson solver
has minor effects on the simulation,
and $10^{-6}$ seems to be small enough.
In the long run,
as the electron bunch expands
more and occupies a bigger region,
PIC must use a large enough domain
to cover all the particles and
avoid non-physical effects caused by
imperfect infinite boundary condition,
so the computation becomes too expensive
and unfeasible.
In addition, it is found that
the particle shape factors matter,
quadratic and cubic shape factors lead
to more accurate results than linear
in the long run.


\section*{Acknowledgment}

The authors acknowledge the support from
National Natural Science Foundation of China
(Grant No. 5247120164).
This research used the open-source particle-in-cell code WarpX
\url{https://github.com/ECP-WarpX/WarpX},
primarily funded by the US DOE Exascale Computing Project.
Primary WarpX contributors are with LBNL, LLNL, CEA-LIDYL,
SLAC, DESY, CERN, and TAE Technologies.
We acknowledge all WarpX contributors.

\section*{Data Availability}
The data that support the findings of this study are available from the corresponding author upon reasonable request.

\bibliographystyle{unsrturl}
\bibliography{reference}

\end{document}